\documentclass{article}
\usepackage{spconf,amsmath,graphicx}
\usepackage{bm}
\usepackage{xcolor}
\usepackage{amsfonts}
\usepackage{amssymb} 
\usepackage{yfonts}
\usepackage{enumitem}
\usepackage{hyperref}
\usepackage{setspace}

\usepackage{tikz}

\usetikzlibrary{patterns}
\usetikzlibrary{backgrounds}
\usetikzlibrary{decorations.text}

\usetikzlibrary{positioning,quotes,calc,fit,shapes,shadows,arrows,trees,mindmap}
\tikzset{block/.style={draw, very thick, minimum height=4cm, align=center}, line/.style={-latex}}
\tikzset{blockV/.style={draw, very thick, text width=2cm, minimum height=2cm, minimum width=4cm, align=center}, line/.style={-latex}}
\tikzset{blockExt/.style={draw, very thick, minimum height=0.7cm, minimum width=0.7cm, align=center}, line/.style={-latex}}

\usepgflibrary{arrows}
\usetikzlibrary{bayesnet}

\definecolor{color_gr}{RGB}{10, 120, 5}
\definecolor{color_gray}{rgb}{0, 0.05, 0.05}
\colorlet{color_vl}{violet!70}
\definecolor{light-gray}{HTML}{E0E0E0}
\definecolor{blue-violet}{rgb}{0.54, 0.17, 0.89}
\definecolor{light-gray}{HTML}{E0E0E0}
\definecolor{carnelian}{rgb}{0.7, 0.11, 0.11}
\definecolor{darkpastelgreen}{rgb}{0.01, 0.75, 0.24}
\usepackage{stmaryrd}
\usepackage{algpseudocode}
\usepackage{algorithm}
\algnewcommand\algorithmicforeach{\textbf{for each}}
\algdef{S}[FOR]{ForEach}[1]{\algorithmicforeach\ #1\ \algorithmicdo}
\usepackage{empheq}
\usepackage{eqparbox}
\newdimen{\algindent}
\usepackage{setspace}
\spacing{1.09}
\algnewcommand\LeftComment[2]{%
\hspace{#1\algindent}$\triangleright$ \eqparbox{COMMENT}{#2} \hfill %
}
\algnewcommand\LeftCommentNoTriangle[2]{%
\hspace{#1\algindent} \eqparbox{COMMENT}{#2} \hfill %
}
\algnewcommand\LeftCommentNoIntent[1]{%
$\triangleright$ \eqparbox{COMMENT}{#1} \hfill %
}
\newcommand{\StateGreen}[1]{\algrenewcommand{\alglinenumber}[1]{\footnotesize\textcolor{darkpastelgreen}{##1}:}\State #1}
\newcommand{\StateBlue}[1]{\algrenewcommand{\alglinenumber}[1]{\footnotesize\textcolor{blue-violet}{##1}:}\State #1}
\newcommand{\StateRed}[1]{\algrenewcommand{\alglinenumber}[1]{\footnotesize\textcolor{carnelian}{##1}:}\State #1}
\newcommand{\StateBlack}[1]{\algrenewcommand{\alglinenumber}[1]{\footnotesize\textcolor{black}{##1}:}\State #1}

\newcommand{\normtwo}[1]{\left\lVert#1\right\rVert_2}

\title{Vector Approximate Message Passing with Arbitrary i.i.d. Noise Priors}
\name{Mohamed Akrout, Tiancheng Gao, Faouzi Bellili, and Amine Mezghani\thanks{This work was supported by the Discovery Grants Program of the Natural Sciences and Engineering Research Council of Canada (NSERC) and Futurewei Technologies.}}
\address{Department of Electrical and Computer Engineering, University of Manitoba, Winnipeg, MB, Canada.}

\begin{document}

\maketitle
\begin{abstract}

Approximate message passing (AMP) algorithms are devised under the Gaussianity assumption of the measurement noise vector. In this work, we relax this assumption within the vector AMP (VAMP) framework to arbitrary independent and identically distributed (i.i.d.) noise priors. We do so by rederiving the linear minimum mean square error (LMMSE) to accommodate both the noise and signal estimations within the message passing steps of VAMP. Numerical results demonstrate how our proposed algorithm handles non-Gaussian noise models as compared to VAMP. This extension to general noise priors enables the use of AMP algorithms in a wider range of engineering applications where non-Gaussian noise models are more appropriate.

\end{abstract}
\begin{keywords}
Approximate message passing, expectation propagation, non-Gaussian noise, inference algorithms.
\end{keywords}
\section{Introduction}
\label{sec:intro}
\subsection{Background and related work}
One of the fundamental problems in the compressed sensing (CS) literature is recovering a sparse vector $\bm{x}\in\mathbb{R}^{N}$ from an observation vector, $\bm{y} \in \mathbb{R}^{M}$, obtained from a noisy linear measurement of the form
\begin{equation}\label{eq:measurement}
    \bm{y}= \bm{A}\,\bm{x}+\bm{w}.
\end{equation}
Here, $\bm{A}\in\mathbb{R}^{M\times N}$ is the measurement matrix with $M < N$, and $\bm{w}$ is unstructured additive noise vector. Solving the inverse problem in (\ref{eq:measurement}) is commonly achieved under the assumption that $\bm{w}$ is an additive white Gaussian noise (AWGN) vector whose entries are assumed to be mutually independent with mean zero and variance $\gamma_w$, i.e., $w_i \sim \mathcal{N}(\mathsf{w}_i; 0, \gamma_w^{-1})$. Since the introduction of the approximate message passing (AMP) algorithm \cite{donoho2009message}, a lot of interest has been devoted to devising algorithmic solutions to (\ref{eq:measurement}) within the CS framework. Specifically, the generalized AMP (GAMP) \cite{rangan2011generalized} algorithm extends over its AMP predecessor by $i)$ accommodating statistical priors on the sparse vector $\bm{x}$ and $ii)$ handling nonlinear output transformations. The divergence issues of the GAMP algorithm due to ill-conditioning of the measurement matrix $\bm{A}$ has been alleviated by the introduction of the vector AMP (VAMP) \cite{DBLP:journals/tit/RanganSF19} \footnote{Note that VAMP was independently derived in \cite{ma2017orthogonal} by another research group under the name orthogonal AMP (OAMP) algorithm.}.

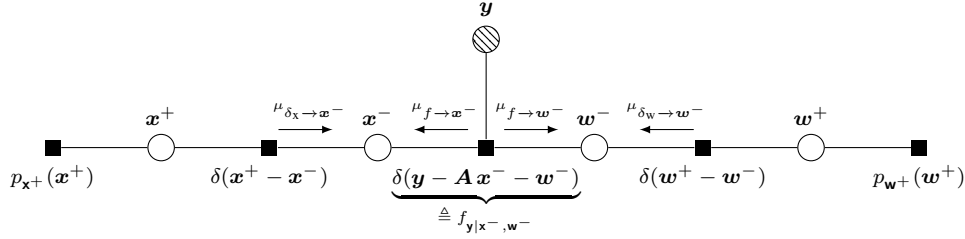
\begin{figure*}[h!]
    \centering
    \begin{tikzpicture}[scale = 1.2]
    \newcommand{\vertex}{\node[vertex]}
    \tikzset{vertex/.style = {circle, draw, inner sep = 0pt, minimum size = 10pt}}
    % variable x+
    \vertex[label = $\bm{x}^{+}$](xp) at (-1.2,0) {};
    % variable x-
    \vertex[label = $\bm{x}^{-}$](xm) at (1.2,0) {};
    % variable w-
    \vertex[label = {$\bm{w}^{-}$}](wm) at (3.6,0) {};
    % variable w-
    \vertex[label = {$\bm{w}^{+}$}](wp) at (6,0) {};
    % variable y
    \vertex[label = $\bm{y}$,pattern = north west lines](y) at (2.4,1.2) {};
    % prior factor x
    \tikzset{vertex/.style = {rectangle, fill = black, inner sep = 0pt, minimum size = 6pt}}
    \vertex[label = below: $p_{\bm{\mathsf{x}}^{+}}(\bm{x}^{+})$](px) at (-2.4,0) {};
    % prior factor delta x
    \vertex[label = below: $\delta(\bm{x}^{+} - \bm{x}^{-})$](delta) at (0,0) {};
    % prior factor y|x,w
    \vertex[label = below: {$\underbrace{\delta(\bm{y} - \bm{A}\,\bm{x}^{-}-\bm{w}^{-})}_{\triangleq~ f_{\bm{\mathsf{y}}|\bm{\mathsf{x}}^{-},\bm{\mathsf{w}}^{-}}}$}](py_xw) at (2.4,0) {}; 
    % prior factor delta w
    \vertex[label = below: {$\delta(\bm{w}^{+} - \bm{w}^{-})$}](deltaw) at (4.8,0) {}; 
    % prior factor w
    \vertex[label = below: $p_{\bm{\mathsf{w}}^{+}}(\bm{w}^{+})$](pw) at (7.2,0) {};
    % arrow
    \draw[-latex] ([xshift=0.2cm,yshift=0.2cm] py_xw.center)--([xshift=-0.4cm,yshift=0.2cm] wm.center);
    \draw[-latex] ([xshift=-0.2cm,yshift=0.2cm] py_xw.center)--([xshift=0.4cm,yshift=0.2cm] xm.center);
    \draw[-latex] ([xshift=-1.1cm,yshift=0.2cm] xm.center)--([xshift=-1.7cm,yshift=0.2cm] py_xw.center);
    \draw[-latex] ([xshift=2.3cm,yshift=0.2cm] py_xw.center)--([xshift=2.9cm,yshift=0.2cm] xm.center);
    % message numbers
    %\node[] at ([xshift=-0.45cm,yshift=0.4cm] py_xw.center) {\circleds{\tiny{$m_1$}}};
    \node[] at ([xshift=-0.45cm,yshift=0.4cm] py_xw.center) {\tiny{$\mu_{f \xrightarrow \,\bm{x}^-}$}};

    \node[] at ([xshift=0.5cm,yshift=0.4cm] py_xw.center) {\tiny{$\mu_{f \xrightarrow \,\bm{w}^-}$}};

    \node[] at ([xshift=-1.95cm,yshift=0.4cm] py_xw.center) {\tiny{$\mu_{\delta_{\text{x}} \xrightarrow \,\bm{x}^-}$}};

    \node[] at ([xshift=2cm,yshift=0.4cm] py_xw.center) {\tiny{$\mu_{\delta_{\text{w}} \xrightarrow \,\bm{w}^-}$}};
    
    % draw lines
    \draw (px)--(xp);
    \draw (delta)--(xp);
    \draw (delta)--(xm);
    \draw (py_xw)--(xm);
    \draw (py_xw)--(wm);
    \draw (py_xw)--(y);
    \draw (deltaw)--(wp);
    \draw (deltaw)--(wm);
    \draw (pw)--(wp);
\end{tikzpicture}
    \vspace{-0.2cm}
    \caption{Factor graph of the VAMP algorithm for arbitrary i.i.d. noise priors. Circles represent variable nodes and  squares represent factor nodes.}
    \vspace{-0.3cm}
    \label{fig:factor-graph}
\end{figure*}

When the noise vector $\bm{w}$ is Gaussian, VAMP strikes a proper balance between the reconstruction performance and the computational complexity as compared to traditional convex optimization-based and iterative soft-thresholding algorithms \cite{eldar2012compressed}. Despite the immense popularity of Gaussian noise models, estimators derived with a Gaussian assumption are sensitive to outliers \cite{kassam2012signal}. In practice, many engineering applications (e.g., electronic devices, lasers, relay switching) and natural phenomena (e.g., atmospheric noise, lightning spikes, and ice cracking) are more accurately characterized by heavy-tailed non-Gaussian measurement noise models \cite{pitas2013nonlinear}. To sidestep the requirement for Gaussian noise models, this paper extends the VAMP algorithm in the presence of arbitrary i.i.d. noise priors, thereby handling a broader class of practical applications.

\subsection{Paper organization and notations}

We structure the rest of this paper as follows. In Section \ref{sec:newAlgorithm}, we present how we incorporate the non-Gaussian noise prior to the expectation propagation steps of the VAMP algorithm. Then, we rederive the new linear minimum mean square error (LMMSE) estimation step of our proposed algorithm to handle arbitrary i.i.d. noise priors. In Section \ref{sec:simulation}, we discuss the simulation results of the algorithm. Finally, we draw out some concluding remarks in Section \ref{Conclusion}.

We also mention the common notations used in this paper. We use Sans Serif fonts (e.g., $\mathsf{x}$) for random variables and Serif fonts (e.g., $x$) for their realizations. We use boldface lowercase letters for vectors (e.g., $\bm{\mathsf{x}}$ and $\boldsymbol{x}$) and we denote the $i$th component of $\boldsymbol{x}$ as $x_i$. The operator diag($\boldsymbol{X}$) stacks the diagonal elements of a matrix $\boldsymbol{X}$ in a vector, $\boldsymbol{I}$ stands for the identity matrix, $\bm{1}$ denotes the all-ones vector. The notation $\bm{\mathsf{x}}\sim$ $p_{\bm{\mathsf{\boldsymbol{x}}}} (\bm{x};\bm{\theta})$ means that $\bm{\mathsf{x}}$ is distributed according to the pdf $p_{\bm{\mathsf{x}}}(\bm{x},\bm{\theta})$ which is parameterized by a parameter vector $\bm{\theta}$. Moreover, $\mathcal{N}(\boldsymbol{x}; \widehat{\bm{x}}, \boldsymbol{R})$ stands for the multivariate Gaussian pdf with mean $\widehat{\bm{x}}$  and covariance matrix $\bm{R}$. We also use $\mathbb{E}[\bm{\mathsf{x}}|d(\boldsymbol{x})]$ and $\text{Cov}[\bm{\mathsf{x}}|d(\boldsymbol{x})]$ to denote the expectation and the covariance of $\bm{{\mathsf{x}}} \sim d(\boldsymbol{x})$. Finally, $\delta(\boldsymbol{x})$ denotes the Dirac delta distribution and $\langle \boldsymbol{x}\rangle \triangleq$ $\frac{1}{N} \sum_{i=1}^{N} x_{i}$ for $\boldsymbol{x} \in \mathbb{R}^{N}$.

\section{VAMP with arbitrary i.i.d. noise priors}
\label{sec:newAlgorithm}

Before delving into the derivation details, we first introduce the VAMP algorithm for arbitrary i.i.d. noise priors which run iteratively according to the algorithmic steps of Algorithm~\ref{algo:algorithm2}. There, $t$ stands for the iteration index, and subscripts $ \mathsf{p}$ and $ \mathsf{e}$ are used to distinguish ``posterior'' and ``extrinsic'' variables, respectively. As a visual reminder, we also use the hat symbol ``$~\,\widehat{}~\,$'' to refer to  mean values.

To derive the VAMP algorithm with arbitrary i.i.d. noise priors, we start by factoring the joint pdf of all the observed and unobserved variables in (\ref{eq:measurement}) as follows:
\begin{equation}\label{eq:joint-pdf}
\begin{aligned}[b]
    p(\boldsymbol{y},\boldsymbol{x}, \boldsymbol{w}) &= p(\boldsymbol{y}|\boldsymbol{x}, \boldsymbol{w})\,p(\boldsymbol{x})\,p(\boldsymbol{w})\\
    &= \delta(\boldsymbol{y}-\bm{A}\,\boldsymbol{x}- \boldsymbol{w})\,p(\boldsymbol{x})\,p(\boldsymbol{w}).
\end{aligned}
\end{equation}
Then, we split both $\boldsymbol{x}$ and $\boldsymbol{w}$ into two identical variables, i.e., $\boldsymbol{x}^+=\boldsymbol{x}^-$ and $\boldsymbol{w}^+=\boldsymbol{w}^-$, thereby transforming the joint pdf in (\ref{eq:joint-pdf}) to the equivalent factorization:
\begin{equation}\label{eq:factorization}
\begin{aligned}[b]
&\small{\!\!p(\boldsymbol{y},\boldsymbol{x}^+,\boldsymbol{x}^-,\boldsymbol{w}^+,\boldsymbol{w}^-)}\\
&\hspace{1cm}\small{~=~ p(\boldsymbol{x}^+)\,p(\boldsymbol{w}^+)\,\delta(\boldsymbol{x}^+-\boldsymbol{x}^-)\times}\\
    &  \hspace{2cm}\small{\delta(\boldsymbol{w}^+-\boldsymbol{w}^-)\,\delta(\boldsymbol{y}-\bm{A}\,\boldsymbol{x}^-- \boldsymbol{w}^-),}
    \end{aligned}
\end{equation}
whose associated factor graph is depicted in Fig. \ref{fig:factor-graph}.

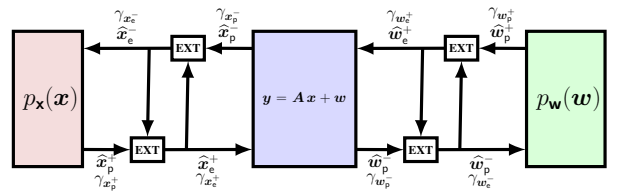
\begin{figure}[h!]
        \centering
         \begin{tikzpicture}[thick,scale=0.45, every node/.style={transform shape}]
  \node[block, fill=carnelian!15] (p_u) {\huge{
  \mbox{$p_{\boldsymbol{\mathsf{x}}}(\boldsymbol{x})$}}};
  \node[block, fill=blue!15, right= 5cm of p_u, minimum width=2.8cm,text width=2.8cm] (z_uv) {\large{$\boldsymbol{y} = \boldsymbol{A}\,\boldsymbol{x}+\boldsymbol{w}$}};
  \node[block, fill=green!15, right= 5cm of z_uv] (phi_z) {\huge{
  \mbox{$p_{\boldsymbol{\mathsf{w}}}(\boldsymbol{w})$}}};
  \node[blockExt,right=of p_u, xshift=0.4cm, yshift=-1.5cm] (ext_pu_to_z_uv) {$\mathrm{\textbf{EXT}}$};
  \node[blockExt,left=of z_uv, xshift=-0.4cm, yshift=1.5cm] (ext_z_uv_to_p_u) {$\mathrm{\textbf{EXT}}$};
  \node[blockExt,right=of z_uv, xshift=0.4cm, yshift=-1.5cm] (ext_z_uv_to_y) {$\mathrm{\textbf{EXT}}$};
  \node[blockExt,left=of phi_z, xshift=-0.4cm, yshift=1.5cm] (ext_y_to_z_uv) {$\mathrm{\textbf{EXT}}$};
  \draw [-latex,very thick] ([yshift=-4.25em]p_u.east) -- 
  node [midway,below=0em,align=center ] { \Large{$\widehat{\boldsymbol{x}}_{\mathsf{p}}^+$}}
  node [midway,below=1.8em,align=center ] {\Large{$\gamma_{\boldsymbol{x}_{\mathsf{p}}^+}$}}
  (ext_pu_to_z_uv.west);
  \draw [-latex,very thick] (ext_pu_to_z_uv) --
  node [midway,below=0.1em,align=center ] { \Large{$\widehat{\boldsymbol{x}}_{\mathsf{e}}^+$}}
  node [midway,below=1.6em,align=center ] {\Large{$\gamma_{\boldsymbol{x}_{\mathsf{e}}^+}$}}
  ([yshift=-4.25em]z_uv.west)
  node [pos=0.25](ext_between_pu_z_uv){};
  % Z=XY to phi(Z)
  \draw [-latex,very thick] ([yshift=-4.25em]z_uv.east) --
  node [midway,below=0em,align=center ] { \Large{$\widehat{\boldsymbol{w}}_{\mathsf{p}}^-$}}
  node [midway,below=1.6em,align=center ] {\Large{$\gamma_{\boldsymbol{w}_{\mathsf{p}}^-}$}}
  (ext_z_uv_to_y.west);
  \draw [-latex,very thick] (ext_z_uv_to_y) --
  node [midway,below=0.1em,align=center ] { \Large{$\widehat{\boldsymbol{w}}_{\mathsf{p}}^-$}}
  node [midway,below=1.7em,align=center ] {\Large{$\gamma_{\boldsymbol{w}_{\mathsf{e}}^-}$}}
  ([yshift=-4.25em]phi_z.west)
  node [pos=0.25](ext_between_z_uv_y){};
  % phi(Z) to Z=XY
  \draw [-latex,very thick] ([yshift=4.25em]phi_z.west) --
  node [midway,above=0em,align=center ] { \Large{$\widehat{\boldsymbol{w}}_{\mathsf{p}}^+$}}
  node [midway,above=1.8em,align=center ] {\Large{$\gamma_{\boldsymbol{w}_{\mathsf{p}}^+}$}}
  (ext_y_to_z_uv.east);
  \draw [-latex,very thick] (ext_y_to_z_uv) --
  node [midway,above=0em,align=center ] { \Large{$\widehat{\boldsymbol{w}}_{\mathsf{e}}^+$}}
  node [midway,above=1.7em,align=center ] {\Large{$\gamma_{\boldsymbol{w}_{\mathsf{e}}^+}$}}
  ([yshift=4.25em]z_uv.east)
  node [pos=0.25](ext_between_y_z_uv){};
  % Z=XY to P(U)
  \draw [-latex,very thick] ([yshift=4.25em]z_uv.west) --
  node [midway,above=0em,align=center ] { \Large{$\widehat{\boldsymbol{x}}_{\mathsf{p}}^-$}}
  node [midway,above=1.7em,align=center ] {\Large{$\gamma_{\boldsymbol{x}_{\mathsf{p}}^-}$}}
  (ext_z_uv_to_p_u.east);
  \draw [-latex,very thick] (ext_z_uv_to_p_u) --
  node [midway,above=0em,align=center ] { \Large{$\widehat{\boldsymbol{x}}_{\mathsf{e}}^-$}}
  node [midway,above=1.7em,align=center ] {\Large{$\gamma_{\boldsymbol{x}_{\mathsf{e}}^-}$}}
  ([yshift=4.25em]p_u.east)
  node [pos=0.25](ext_between_z_uv_pu){};
  %  Z=XY to P(V)
  %\draw [-latex,very thick] ([xshift=1.75em]z_uv.north) --
  %node [pos=0.6,right=0em,align=center ] { $\boldsymbol{\hat{V}}_{\alpha}^{-}$}
  %node [pos=0.2,right=0em,align=center ] {$\boldsymbol{\Lambda}_{\boldsymbol{V}_{\alpha}^{-}}$}
  %(ext_z_uv_to_v.south);
  %\draw [-latex,very thick] (ext_z_uv_to_v.north) --
  %node [pos=0.6,right=0em,align=center ] { $\boldsymbol{\hat{V}}_{\eta}^{-}$}
  %node [pos=0.4,right=0em,align=center ] {$\gamma_{\boldsymbol{V}^{-}_{\eta}}$}
  %([xshift=1.75em]p_v.south)
  %node [pos=0.25](ext_between_z_uv_pv){};
  % P(V) to Z=XV
  %\draw [-latex,very thick] ([xshift=-1.75em]p_v.south) --
  %node [pos=0.6,left=0em,align=center ] { $\boldsymbol{\hat{V}}_{\alpha}^{+}$}
  %node [pos=0.25,left=0em,align=center ] {$\gamma_{\boldsymbol{V}^{+}_{\alpha}}$}
  %(ext_v_to_z_uv.north);
  %\draw [-latex,very thick] (ext_v_to_z_uv.south) --
  %node [pos=0.6,left=0em,align=center ] { $\boldsymbol{\hat{V}}_{\eta}^{+}$}
  %node [pos=0.4,left=0em,align=center ] {$\gamma_{\boldsymbol{V}^{+}_{\eta}}$}
  %([xshift=-1.75em]z_uv.north)
  %node [pos=0.25](ext_between_pv_z_uv){};
  
  % Extrinsic Arrows
  \draw [-latex,very thick] (ext_between_pu_z_uv.center) --
  (ext_z_uv_to_p_u.south);
  \draw [-latex,very thick] (ext_between_z_uv_y.center) --
  (ext_y_to_z_uv.south);
  \draw [-latex,very thick] (ext_between_y_z_uv.center) --
  (ext_z_uv_to_y.north);
  \draw [-latex,very thick] (ext_between_z_uv_pu.center) --
  (ext_pu_to_z_uv.north);
  %\draw [-latex,very thick] (ext_between_z_uv_pv.center) --
  %(ext_v_to_z_uv.east);
  %\draw [-latex,very thick] (ext_between_pv_z_uv.center) --
  %(ext_z_uv_to_v.west);
\end{tikzpicture}
         \vspace{-0.1cm}
        \caption{Block diagram of VAMP for arbitrary i.i.d. noise priors with its three modules: two denoising MMSE modules incorporating the prior information, $p_{\bm{\mathsf{x}}}(\cdot)$ and $p_{\bm{\mathsf{w}}}(\cdot)$, and the LMMSE module. The three modules exchange extrinsic information/messages through the \protect\tikz[inner sep=.25ex,baseline=-.75ex] \protect\node[rectangle,draw,thick,minimum width=0.45cm,minimum height=0.45cm] {\footnotesize \textbf{ext}}; blocks. The color of each module matches the color of the corresponding line numbers in Algorithm~\ref{algo:algorithm2}.}
        \label{fig:block-diagram}
\end{figure}

\noindent Similarly to the derivation of standard VAMP, we follow an expectation propagation (EP)-like approximation of the sum-product (SP) belief propagation (BP) algorithm based on the two following rules: 
\begin{itemize}[leftmargin=*]
    \item \underline{\textit{EP approximation}}: Given a variable node $\bm{\mathsf{x}}^-$ receiving the message  $\mu_{f\rightarrow \bm{\mathsf{x}}^-}$ from the factor node $f_{\bm{\mathsf{y}}|\bm{\mathsf{x}}^{-},\bm{\mathsf{w}}^{-}} = \delta(\bm{y} - \bm{A}\,\bm{x}^{-}-\bm{w}^{-})$, EP approximates the SP belief $b_{\text{sp}}(\boldsymbol{x}^-)$ with the approximate belief $b_{\text{app}}(\boldsymbol{x}^{-})= \mathcal{N}(\bm{x}^-; \widehat{\boldsymbol{x}}_{\mathsf{p}}^-,\gamma_{\boldsymbol{x}_{\mathsf{p}}^-}^{-1})$, where $\widehat{\boldsymbol{x}}_{\mathsf{p}}^- = \mathbb{E}[\boldsymbol{x}^-|b_{\text{sp}}]$ and $\gamma_{\boldsymbol{x}_{\mathsf{p}}^-}^{-1} = \left\langle\operatorname{diag}\left(\operatorname{Cov}\left[\boldsymbol{x}^- | b_{\text{sp}}\right]\right)\right\rangle$ are the mean and average variance of the SP belief $b_{\text{sp}}(\boldsymbol{x}^-)$.
    \item \underline{\textit{Extrinsic belief computation}}: As shown in Fig~\ref{fig:block-diagram}, given a posterior Gaussian belief $\mathcal{N}(\bm{x}^-; \widehat{\boldsymbol{x}}_{\mathsf{p}}^-,\gamma_{\boldsymbol{x}_{\mathsf{p}}^-}^{-1}\, \boldsymbol{I})$ and an incoming extrinsic Gaussian belief $\mathcal{N}(\bm{x}^-; \widehat{\boldsymbol{x}}_{\mathsf{e}}^+,\gamma_{\boldsymbol{x}_{\mathsf{e}}^+}^{-1}\, \boldsymbol{I})$, the extrinsic mean and precision of the outcoming Gaussian belief $\mathcal{N}(\bm{x}^+; \widehat{\boldsymbol{x}}_{\mathsf{e}}^-,\gamma_{\boldsymbol{x}_{\mathsf{e}}^-}^{-1}\, \boldsymbol{I})$ are computed as follows:\vspace{-0.12cm}
\begin{subequations}\label{eq:extrinsic-computation}
        \begin{align}
        \!\!\!\!\gamma_{\boldsymbol{x}_{\mathsf{e}}^-} &\,\,= \,\, \gamma_{\boldsymbol{x}_{\mathsf{p}}^-} - \gamma_{\boldsymbol{x}_{\mathsf{e}}^+} ~\triangleq~ \text{EXT}\left(\gamma_{\boldsymbol{x}_{\mathsf{p}}^-}\right),\label{eq:extrinsic-precision}\\
        \!\!\!\!\widehat{\boldsymbol{x}}_{\mathsf{e}}^-  &\,\,=\,\, \gamma_{\boldsymbol{x}_{\mathsf{e}}^+}^{-1}\left(\gamma_{\boldsymbol{x}_{\mathsf{p}}^-} \,\widehat{\boldsymbol{x}}_{\mathsf{p}}^-- \gamma_{\boldsymbol{x}_{\mathsf{e}}^+}\,\widehat{\boldsymbol{x}}_{\mathsf{e}}^+\right) ~\triangleq~\text{EXT}\left(\widehat{\boldsymbol{x}}_{\mathsf{p}}^-\right). \label{eq:extrinsic-mean}
        \end{align}
    \end{subequations}
\end{itemize}

\noindent Unlike the factor graph of standard VAMP, however, the two variable nodes $\boldsymbol{x}^-$ and $\boldsymbol{w}^-$ in Fig.~\ref{fig:factor-graph} are connected through the factor node $f_{\bm{\mathsf{y}}|\bm{\mathsf{x}}^{-},\bm{\mathsf{w}}^{-}} = \delta(\bm{y} - \bm{A}\,\bm{x}^{-}-\bm{w}^{-})$ stemming from the measurement model in (\ref{eq:measurement}). This means that the LMMSE estimate of $\bm{\mathsf{x}}^-$ must account for the estimate of $\bm{\mathsf{x}}^-$ and vice versa. Note that the denoising steps corresponding to the red and green modules in Fig. \ref{fig:block-diagram} are similar to the denoising procedure in VAMP. Thus, to customize the original VAMP framework so as to recover the signal $\boldsymbol{x}$ under an arbitrary noise prior $p_{\bm{\mathsf{w}}}(\bm{w})$, one should derive from scratch the LMMSE step for both $\bm{\mathsf{x}}^-$ and $\bm{\mathsf{w}}^-$.\vspace{-0.2cm}

\begin{algorithm}[H]
\small
\caption{VAMP with arbitrary i.i.d. noise priors}\label{algo:algorithm2}
\begin{algorithmic}[1]
\Statex $\mathbf{Require:}$ The channel matrix $\boldsymbol{A}\in \mathbb{R}^{M \times N}$; the received vector $\boldsymbol{y}\in\mathbb{R}^{M}$; the maximum number of iterations $T_{\text{max}}$.
\State $\mathbf{Initialize\,:\,}~\widehat{\boldsymbol{x}}^-_{\mathsf{e},0},\gamma_{\boldsymbol{x}^{-}_{\mathsf{e}},0}$, $\widehat{\boldsymbol{w}}^-_{\mathsf{e},0}$ and $\gamma_{\boldsymbol{w}^{-}_{\mathsf{e}},0}$ \vspace{0.08cm}

\For {$t=0,\dots, T_{\text{max}}-1$}\vspace{0.1cm}
\Statex \LeftComment{1}{\textcolor{carnelian}{\underline{Denoising $\boldsymbol{x}$}}}\vspace{0.1cm}
\StateRed$\boldsymbol{\widehat{x}}^+_{\mathsf{p},t}\;=\; g_{\bm{\mathsf{x}}}\left(\widehat{\boldsymbol{x}}^-_{\mathsf{e},t}, \gamma_{\boldsymbol{x}^{-}_{\mathsf{e}},t}\right)$\label{algo:line3}\vspace{0.03cm} %\mathbb{E}[x_{1}|b(x_{1})]$ %\hspace{1cm}// $b(x_{1})$ represent the belief on $\boldsymbol{x}_{1}$
\State  $\alpha_{\boldsymbol{x}_{\mathsf{p}}^+,t} = \frac{1}{L}\,\sum\limits_{j=1}^{L}\!g_{\bm{\mathsf{x}}}^{\prime}\left(\widehat{x}^-_{j,\mathsf{e},t}, \gamma_{\boldsymbol{x}^{-}_{\mathsf{e}},t}\right)$  \label{algo:line4} \vspace{0.03cm}
\State $\gamma_{\boldsymbol{x}_{\mathsf{p}}^+,t}=\gamma_{\boldsymbol{x}_{\mathsf{e}}^-,t}/\alpha_{\boldsymbol{x}_{\mathsf{p}}^+,t}$\label{algo:line5}\vspace{0.18cm}
\StateBlack $\gamma_{\boldsymbol{x}_{\mathsf{e}}^+,t}\;=\; \gamma_{\boldsymbol{x}_{\mathsf{p}}^+,t}-\gamma_{\boldsymbol{x}_{\mathsf{e}}^-,t}$\label{algo:line6}\vspace{0.18cm}
\State $\boldsymbol{\widehat{x}}^+_{\mathsf{e},t}\;=\; (\gamma_{\boldsymbol{x}_{\mathsf{p}}^+,t}\boldsymbol{\widehat{x}}^+_{\mathsf{p},t}-\gamma_{\boldsymbol{x}^{-}_{\mathsf{e}},t}\widehat{\boldsymbol{x}}^-_{\mathsf{e},t})/\gamma_{\boldsymbol{x}_{\mathsf{e}}^+,t}$\label{algo:line7}\vspace{0.08cm}
\Statex\LeftComment{1}{\textcolor{darkpastelgreen}{\underline{Denoising $\boldsymbol{w}$}}}\vspace{0.13cm}
\StateGreen $\boldsymbol{\widehat{w}}^+_{\mathsf{p},t}\;=\; g_{\bm{\mathsf{w}}}\left(\widehat{\boldsymbol{w}}^-_{\mathsf{e},t}, \gamma_{\boldsymbol{w}^{-}_{\mathsf{e}},t}\right)$ \label{algo:line8}\vspace{0.05cm}
\State $\alpha_{\boldsymbol{w}_{\mathsf{p}}^+,t} = \frac{1}{L}\,\sum\limits_{j=1}^{L} g_{\bm{\mathsf{w}}}^{\prime}\left(\widehat{w}^-_{j,\mathsf{e},t}, \gamma_{\boldsymbol{w}^{-}_{\mathsf{e}},t}\right)$\label{algo:line9}\vspace{0.09cm}
\State $\gamma_{\boldsymbol{w}_{\mathsf{p}}^+,t}=\gamma_{\boldsymbol{w}_{\mathsf{e}}^-,t}/\alpha_{\boldsymbol{w}_{\mathsf{p}}^+,t}$\label{algo:line10}\vspace{0.18cm}
\StateBlack $\gamma_{\boldsymbol{w}_{\mathsf{e}}^+,t}\;=\; \gamma_{\boldsymbol{w}_{\mathsf{p}}^+,t}-\gamma_{\boldsymbol{w}_{\mathsf{e}}^-,t} $\label{algo:line11}\vspace{0.18cm}
\State $\boldsymbol{\widehat{w}}^+_{\mathsf{e},t}\;=\; (\gamma_{\boldsymbol{w}_{\mathsf{p}}^+,t}\,\boldsymbol{\widehat{w}}^+_{\mathsf{p},t}-\gamma_{\boldsymbol{w}^{-}_{\mathsf{e}},t}\,\widehat{\boldsymbol{w}}^-_{\mathsf{e},t})/\gamma_{\boldsymbol{w}_{\mathsf{e}}^+,t}$\label{algo:line12}\vspace{0.08cm}

\Statex\LeftComment{1}{\textcolor{blue-violet}{\underline{LMMSE estimation of $\boldsymbol{x}$ and $\boldsymbol{w}$}}}\vspace{0.15cm}
\StateBlue $\boldsymbol{\widehat{x}}^-_{\mathsf{p},t} = \textswab{f}\big(\boldsymbol{x}_{\mathsf{e},t}^+, \gamma_{\boldsymbol{x}_{\mathsf{e},t}^+},\boldsymbol{w}_{\mathsf{e},t}^+, \gamma_{\boldsymbol{w}_{\mathsf{e},t}^+}\big)$\label{algo:line13}
\State $\boldsymbol{\widehat{w}}^-_{\mathsf{p},t} = \textswab{g}\big(\boldsymbol{x}_{\mathsf{e},t}^+, \gamma_{\boldsymbol{x}_{\mathsf{e},t}^+},\boldsymbol{w}_{\mathsf{e},t}^+, \gamma_{\boldsymbol{w}_{\mathsf{e},t}^+}\big)$\label{algo:line13_1}
%\StateBlue  $\boldsymbol{\widehat{x}}^-_{\mathsf{p},t},\boldsymbol{\widehat{z}}^-_{\mathsf{p},t}=g_{2}\left(\boldsymbol{\widehat{x}}^+_{\mathsf{e},t}, \gamma_{\boldsymbol{x}_{\mathsf{e}}^+,t},\boldsymbol{\widehat{z}}^+_{\mathsf{e},t}, \gamma_{\boldsymbol{z}_{\mathsf{e}}^+,t}\right)$\label{algo:line13}\vspace{0.08cm}  %// $b(\boldsymbol{x}_{2},\boldsymbol{z}_{2})$ represents\vspace{0.05cm} the belief on the extended LMMSE block $\{\boldsymbol{x}_{2},\boldsymbol{z}_{2},\delta\big(g(\boldsymbol{x}_{2},\boldsymbol{z}_{2})=0\big)\}$\vspace{0.07cm}

\State $\gamma_{\boldsymbol{x}_{\mathsf{p}}^-,t}\;=\; \gamma_{\boldsymbol{x}_{\mathsf{e}}^+,t}/\alpha_{\boldsymbol{x}_{\mathsf{p}}^-,t}$\label{algo:line14}~\,~\, with \,~\, $\alpha_{\boldsymbol{x}_{\mathsf{p}}^-,t}\;=\;\left\langle \textswab{f}^{\prime}\right\rangle$\label{algo:line14}\vspace{-0.01cm}
%\State $\alpha_{\boldsymbol{z},2k}\;=\;\left\langle q^{\prime}\right\rangle$\label{algo:line15}\vspace{0.11cm}
%\State ......\vspace{0.2cm}
\State $\gamma_{\boldsymbol{w}_{\mathsf{p}}^-,t}\;=\; \gamma_{\boldsymbol{w}_{\mathsf{e}}^+,t}/\alpha_{\boldsymbol{w}_{\mathsf{p}}^-,t}$\,~\,~\, with\,~\, ~$\alpha_{\boldsymbol{w}_{\mathsf{p}}^-,t}\;=\;\left\langle \textswab{g}^{\prime}\right\rangle$ \label{algo:line19}\vspace{0.18cm}
\State $\gamma_{\boldsymbol{x}_{\mathsf{e}}^-,t+1}\;=\; \gamma_{\boldsymbol{x}_{\mathsf{p}}^-,t}-\gamma_{\boldsymbol{x}_{\mathsf{e}}^+,t}$\label{algo:line17}\vspace{0.18cm}
\State $\gamma_{\boldsymbol{w}_{\mathsf{e}}^-,t+1}\;=\gamma_{\boldsymbol{w}_{\mathsf{p}}^-,t}-\gamma_{\boldsymbol{w}_{\mathsf{e}}^+,t}$ 
\State $\widehat{\boldsymbol{x}}^-_{\mathsf{e},t+1}\;=\; \left(\gamma_{\boldsymbol{x}_{\mathsf{p}}^-,t}\,\boldsymbol{\widehat{x}}^-_{\mathsf{p},t}-\gamma_{\boldsymbol{x}^{+}_{\mathsf{e}},t}\,\widehat{\boldsymbol{x}}^+_{\mathsf{e},t}\right)\Big/\gamma_{\boldsymbol{x}_{\mathsf{e}}^+,t+1}$\label{algo:line18}\vspace{0.18cm}
%\State $\gamma_{\boldsymbol{z},1\,k+1}\;=\; \eta_{\boldsymbol{z},2k}-\gamma_{\boldsymbol{z},2k}$\label{algo:line20}\vspace{0.2cm}
\State $\widehat{\boldsymbol{w}}^-_{\mathsf{e},t+1}\;=\; \left(\gamma_{\boldsymbol{w}_{\mathsf{p}}^-,t}\boldsymbol{\widehat{w}}^-_{\mathsf{p},t}-\gamma_{\boldsymbol{w}^{+}_{\mathsf{e}},t}\widehat{\boldsymbol{w}}^+_{\mathsf{e},t}\right)\Big/\gamma_{\boldsymbol{w}_{\mathsf{e}}^+,t+1}$\label{algo:line21}\vspace{0.06cm}
\EndFor
\vspace{0.015cm}
\State \textbf{Return} ~$\widehat{\boldsymbol{x}}^+_{\mathsf{p},t}$
\end{algorithmic}
\end{algorithm}

To derive the joint LMMSE estimates $\boldsymbol{\widehat{x}}^-_{\mathsf{p}}$ and $\boldsymbol{\widehat{w}}^-_{\mathsf{p}}$ of $\bm{\mathsf{x}}^-$ and $\bm{\mathsf{w}}^-$, we write the joint belief of $\boldsymbol{x}^-$ and $\boldsymbol{w}^-$ pertaining to their joint pdf in (\ref{eq:joint-pdf}) as the product of the three beliefs depicted in Fig. \ref{fig:factor-graph}:
\begin{equation}\label{eq:g2-joint-belief}
\begin{aligned}[b]
    b(\boldsymbol{x}^-,\boldsymbol{w}^-) &\propto \mu_{\delta_{\text{x}} \rightarrow \boldsymbol{x}^- }(\boldsymbol{x}^-)\,\cdot\,f_{\bm{\mathsf{y}}|\bm{\mathsf{x}}^{-},\bm{\mathsf{w}}^{-}}\,\cdot\, \mu_{\delta_{\text{w}} \rightarrow \boldsymbol{w}^- }(\boldsymbol{w}^-)\\
    &=\mathcal{N}(\boldsymbol{x}^-;\widehat{\boldsymbol{x}}_{\mathsf{e}}^+,\gamma_{\boldsymbol{x}_{\mathsf{e}}^+}^{-1}\, \mathbf{I}_N)~\delta(\bm{y} - \bm{A}\,\bm{x}^{-}-\bm{w}^{-})\\
    &~~~~~~~~~~\hspace{1cm}\times\mathcal{N}(\boldsymbol{w}^-;\widehat{\boldsymbol{w}}_{\mathsf{e}}^+, \gamma_{\boldsymbol{w}_{\mathsf{e}}^+}^{-1}\mathbf{I}_M).
    \end{aligned}
\end{equation}

\noindent Finding the LMMSE estimates $\boldsymbol{\widehat{x}}^-_{\mathsf{p}}$ and $\boldsymbol{\widehat{w}}^-_{\mathsf{p}}$ boils down to evaluating the following integrals:
\begin{subequations}\label{eq:g2-MMSE-integrals}
\begin{align}
    \boldsymbol{\widehat{x}}^-_{\mathsf{p}} &=\frac{ \int \int\boldsymbol{x}^-\,b(\boldsymbol{x}^-,\boldsymbol{w}^-)\, \text{d}\boldsymbol{x}^-\,\text{d}\boldsymbol{w}^-}{\int\int\,b(\boldsymbol{x}^-,\boldsymbol{w}^-)\,\text{d}\boldsymbol{x}^-\,\text{d}\boldsymbol{w}^-}\nonumber\\
    &= \left(\gamma_{\boldsymbol{x}_{\mathsf{e}}^+}\,\mathbf{I}_N + \gamma_{\boldsymbol{w}_{\mathsf{e}}^+}\bm{A}^{\top}\bm{A}\right)^{-1}\nonumber\\
    &\hspace{0.8cm}\times \left(\gamma_{\boldsymbol{x}_{\mathsf{e}}^+}\,\boldsymbol{\widehat{x}}_{\mathsf{e}}^+ + \gamma_{\boldsymbol{w}_{\mathsf{e}}^+}\,\bm{A}^{\top}\left(\bm{y}-\boldsymbol{\widehat{w}}_{\mathsf{e}}^+\right)\right)\nonumber\\
    &\triangleq \textswab{f}\left(\boldsymbol{\widehat{x}}_{\mathsf{e}}^+, \gamma_{\boldsymbol{x}_{\mathsf{e}}^+},\boldsymbol{\widehat{w}}_{\mathsf{e}}^+, \gamma_{\boldsymbol{w}_{\mathsf{e}}^+}\right),\\ 
    \boldsymbol{\widehat{w}}^-_{\mathsf{p}} &=\frac{ \int \int \boldsymbol{w}^-\,b(\boldsymbol{x}^-,\boldsymbol{w}^-)\, \text{d}\boldsymbol{w}^-\,\text{d}\boldsymbol{x}^-}{\int\int\,b(\boldsymbol{x}^-,\boldsymbol{w}^-)\,\text{d}\boldsymbol{w}^-\text{d}\boldsymbol{x}^-}\nonumber\\
    &= \left(\gamma_{\boldsymbol{w}_{\mathsf{e}}^+}\,\mathbf{I}_M + \gamma_{\boldsymbol{x}_{\mathsf{e}}^+}\bm{Q}\right)^{-1}\nonumber\\
    &\hspace{0.8cm}\times \left(\gamma_{\boldsymbol{w}_{\mathsf{e}}^+}\,\boldsymbol{\widehat{w}}_{\mathsf{e}}^+ + \gamma_{\boldsymbol{x}_{\mathsf{e}}^+}\,\bm{Q}\left(\bm{y}-\bm{A}\,\boldsymbol{\widehat{x}}_{\mathsf{e}}^+\right)\right)\nonumber\\
    &\triangleq \textswab{g}\left(\boldsymbol{\widehat{x}}_{\mathsf{e}}^+, \gamma_{\boldsymbol{x}_{\mathsf{e}}^+},\boldsymbol{\widehat{w}}_{\mathsf{e}}^+, \gamma_{\boldsymbol{w}_{\mathsf{e}}^+}\right).
\end{align}
\end{subequations}
where $\bm{Q} \triangleq \left(\bm{A}\,\bm{A}^{\top}\right)^{-1}$. Finally, to find the posterior precisions $\gamma_{\boldsymbol{x}_{\mathsf{p}}^-}$ and $\gamma_{\boldsymbol{w}_{\mathsf{p}}^-}$ of $\boldsymbol{x}^-$ and $\boldsymbol{w}^-$, we define the divergences of $\textswab{f}(\cdot)$ and $\textswab{g}(\cdot)$ at $\boldsymbol{x}_{\mathsf{e}}^+$ and $\boldsymbol{w}_{\mathsf{e}}^+$ as follows \cite{DBLP:journals/tit/RanganSF19}:

\begin{subequations}
    \begin{align}
    \textswab{f}^{\prime} &= \text{diag}\left(\frac{\partial \,\textswab{f}}{\partial\, \widehat{\boldsymbol{x}}_{\mathsf{e}}^+}\right) =\gamma_{\boldsymbol{x}_{\mathsf{e}}^+}\,\text{diag}\small{\left( \left(\gamma_{\boldsymbol{x}_{\mathsf{e}}^+}\,\mathbf{I}_N + \gamma_{\boldsymbol{w}_{\mathsf{e}}^+}\bm{A}^{\top}\bm{A}\right)^{-1}\right)},\\
    \textswab{g}^{\prime} &=\text{diag}\left(\frac{\partial \,\textswab{g}}{\partial\, \widehat{\boldsymbol{w}}_{\mathsf{e}}^+}\right) = \gamma_{\boldsymbol{w}_{\mathsf{e}}^+}\,\text{diag}\left(\left(\gamma_{\boldsymbol{w}_{\mathsf{e}}^+}\,\mathbf{I}_M + \gamma_{\boldsymbol{x}_{\mathsf{e}}^+}\bm{Q}\right)^{-1} \right).
    \end{align}
\end{subequations}

\noindent 
Now, after defining
 $\alpha_{\boldsymbol{x}^-}\triangleq\left\langle \textswab{f}^{\prime}\right\rangle$ and  $\alpha_{\boldsymbol{w}^-}\triangleq\left\langle \textswab{g}^{\prime}\right\rangle$, we use the fact that \cite{DBLP:journals/tit/RanganSF19}
\begin{equation}\label{eq:alphas-x-z}
\begin{aligned}
\left\langle \textswab{f}^{\prime}\right\rangle=\frac{\gamma_{\boldsymbol{x}_{\mathsf{e}}^+}}{\gamma_{\boldsymbol{x}_{\mathsf{p}}^-}},~\text{and}~~
\left\langle \textswab{g}^{\prime}\right\rangle=\frac{\gamma_{\boldsymbol{z}_{\mathsf{e}}^+}}{\gamma_{\boldsymbol{z}_{\mathsf{p}}^-}},
\end{aligned} 
\end{equation}

\noindent to deduce the posterior precisions $\gamma_{\boldsymbol{x}_{\mathsf{p}}^-}$ and $\gamma_{\boldsymbol{w}_{\mathsf{p}}^-}$ as shown in lines \ref{algo:line14}--\ref{algo:line19} of Algorithm \ref{algo:algorithm2}.

\section{Simulation Results}
\label{sec:simulation}

In this section, we assess the estimation performance of the proposed VAMP algorithm with i.i.d. priors and benchmark it against the standard VAMP algorithm. In all simulations, we set the number of time steps to $T_{\textrm{max}}=100$, and perform $N_{\textrm{MC}}=100$ Monte-Carlo trials for different values of the SNR defined in dB as
\begin{equation}
\label{eq:SNR}
     \textrm{SNR} \,=\, \textrm{10 log}_{\textrm{10}}\Bigg(\frac{ \normtwo{\bm{A}\,\bm{x}}^{2}}{\normtwo{\boldsymbol{w}}^{2}}\Bigg).
\end{equation}

\noindent Here, each element $a_{ij}$ of $\bm{A}$ is drawn from the standard Gaussian distribution \big(i.e., $a_{ij} \sim \mathcal{N}(0,1)$\big), and the vector  $\bm{x}_{\ell}$ is drawn from a Bernoulli-Gaussian density:
\begin{equation}
\label{eq:innovation-prior-BG}
    p_{\bm{\mathsf{x}}}(\bm{x}) = \rho\,\delta(\bm{x}) + (1-\rho)\,\mathcal{N}\left(\bm{x}; \bm{0},\mathbf{I}_N\right),
\end{equation}
where $\rho$ is the sparsity rate (i.e., the percentage of the non-zero components). We use the normalized root MSE (NRMSE) as a performance measure which is defined as:
\begin{equation*}
    \textrm{NRMSE} ~=~\frac{1}{N_{\textrm{MC}}}\sum_{\ell=1}^{N_{\textrm{MC}}} \frac{\|\bm{A}\,\bm{x}_{\ell}-\bm{A}\,\widehat{\bm{x}}_{\ell}\|_{2}}{\|\bm{A}\,\bm{x}_{\ell}\|_{2}}
\end{equation*}
where $\bm{x}_{\ell}$ is $\ell$th realization of $\bm{\mathsf{x}}$ and $\widehat{\bm{x}}_{\ell}$ is its reconstruction during the $\ell$th Monte-Carlo trial. 

\noindent We examine the NRMSE for two i.i.d. non-Gaussian priors modelling each component $w_i$ of the noise vector $\bm{w}$ as a realization draw from
\begin{itemize}[leftmargin=*]
    \item the \textit{Laplace prior} $\,p_{{\mathsf{w}}}(w_i;\mu,b) = \frac{1}{2b}\,\textrm{exp}\left(-\frac{|w
    _i-\mu|}{b}\right)$ which better models heavy-tailed non-Gaussian noise sources (e.g., in electronic devices).
    
    \item the \textit{binary prior} $\,p_{{\mathsf{w}}}(w_i; s)=\frac{1}{2}\,\delta(w_i-s) + \frac{1}{2}\,\delta(w_i+s)$ which can represent erasing noise sources or bit-flip distortions and corruptions (e.g., in printing process).
\end{itemize}

\begin{figure}[h!]
\begin{minipage}[b]{\linewidth}
  \centering
  \centerline{\includegraphics[scale=0.3]{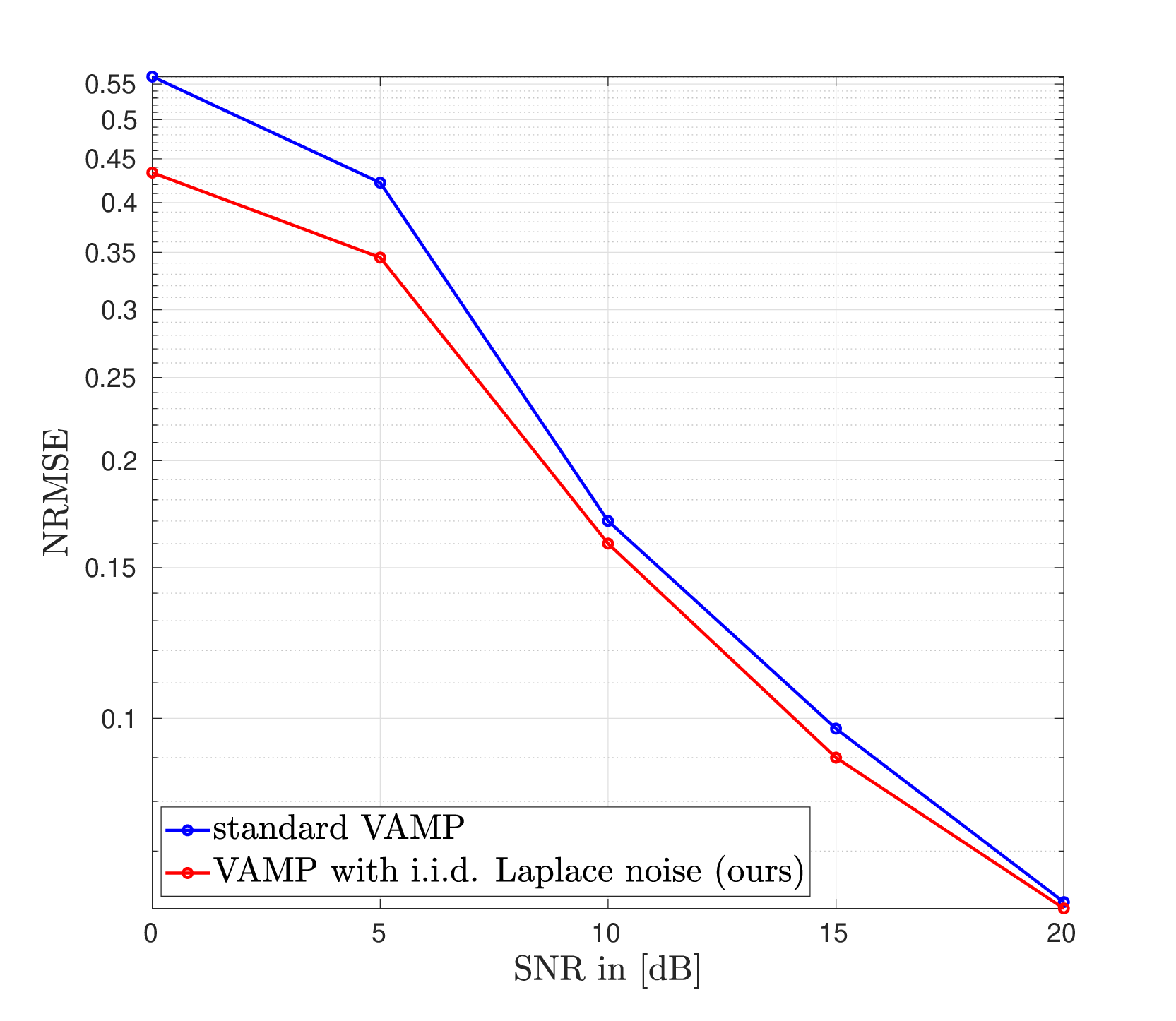}}
\end{minipage}
\vspace{-0.6cm}
\caption{NRMSE of VAMP with arbitrary i.i.d. priors vs. standard VAMP as a function of the SNR with the noise vector drawn from the Laplace distribution with $\mu=0$.\vspace{-0.15cm}}
\label{fig:NRMSE-Laplace}
\end{figure}

\noindent We start by comparing standard VAMP to our algorithm when the noise vector $w$ is drawn from the Laplace distribution $\mu=0$ while varying $b$ to meet the SNR level defined in (\ref{eq:SNR}). We also set $\rho = 95\%$. Fig. \ref{fig:NRMSE-Laplace} depicts the NRMSE of both algorithms and shows how incorporating the noise model improves the reconstruction accuracy, especially in the low SNR region. The fact that Laplace and Gaussian distributions are not very far (in the KL-divergence sense) diminishes the improvement of our algorithm in the high SNR regime since both distributions are centered around the same mean $\mu = 0$.

\begin{figure}[h!]
\begin{minipage}[b]{\linewidth}
  \centering
  \centerline{\includegraphics[scale=0.3]{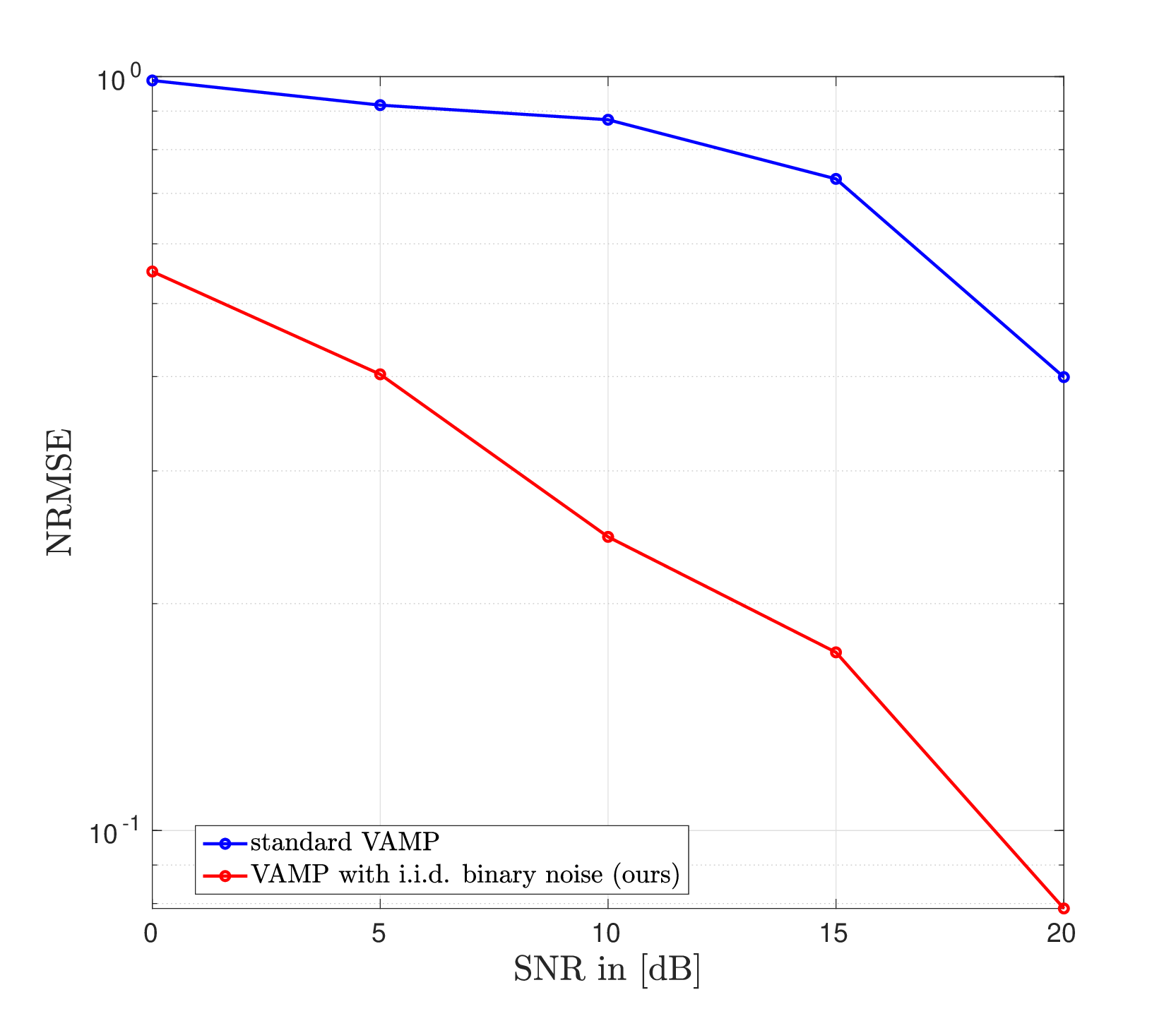}}
\end{minipage}
\vspace{-0.6cm}
\caption{NRMSE of VAMP with arbitrary i.i.d. priors vs. standard VAMP as a function of the SNR with the noise vector drawn from the binary distribution.\vspace{-0.15cm}}
\label{fig:NRMSE-binary}
\end{figure}

\noindent Significant improvements can be observed as soon as the noise considerably deviates from the Gaussian model. When the noise vector $\bm{w}$ is drawn from the binary prior with $a=1$, our algorithm reveals more robust to discrete noise sources and outperforms the standard VAMP over the entire SNR range [0 dB, 20 dB] as depicted in Fig. \ref{fig:NRMSE-binary}. There, unlike our observation in Fig. \ref{fig:NRMSE-Laplace}, it is seen how the improvement of our algorithm increases as the SNR increases because of the substantial mismatch between the binary and Gaussian distributions.

\section{Conclusion}
\label{Conclusion}

In this paper, we build upon the vector approximate message passing algorithm to handle non-Gaussian measurement noise models. We did so by incorporating the arbitrary noise prior into standard VAMP by solving a joint LMMSE estimation problem in addition to the MMSE noise denoiser developed in the paper. Computer simulations with Laplace and binary noise models confirmed that our proposed algorithm exhibits significant reconstruction improvements as a function of the mismatch between the considered noise distribution and the Gaussian one.

\bibliographystyle{IEEEbib}
\bibliography{refs}

\end{document}